\documentclass[twocolumn,superscriptaddress,prl,aps,10pt]{revtex4-1}

\usepackage{amsfonts,amssymb}
\usepackage[sumlimits,intlimits]{amsmath}
\usepackage{graphics}
\usepackage{epsfig}

\usepackage{mathrsfs}
\usepackage{textcomp}
\usepackage{verbatim}
\usepackage{bm}          
\usepackage{enumerate}
\usepackage{multirow}
\usepackage{dcolumn}

\usepackage[colorlinks = true,
            linkcolor = blue,
            urlcolor  = blue,
            citecolor = blue,
            anchorcolor = blue,
            bookmarks=false]{hyperref}

\newcommand{\figref}[1]{Fig.~\ref{#1}}

\newcommand{\be}{\begin{equation}}
\newcommand{\ee}{\end{equation}}

\newcommand{\eq}[1]{Eq.\ (\ref{eq:#1})}

\usepackage{xspace}

\newcommand{\dcomment}[1]{\color{blue}{#1}\color{black}\xspace}

\newcommand{\IGNORE}[1]{}

\graphicspath{{./figures/}}

\begin{document}

\title{Universal power law governing pedestrian interactions}

\author{Ioannis Karamouzas}
\affiliation{Department of Computer Science and Engineering, University of Minnesota, USA}
\author{Brian Skinner}
\affiliation{Materials Science Division, Argonne National Laboratory, USA}
\author{Stephen J. Guy}
\affiliation{Department of Computer Science and Engineering, University of Minnesota, USA}

\date{\today}

\begin{abstract}
Human crowds often bear a striking resemblance to interacting particle systems, and this has prompted many researchers to describe pedestrian dynamics in terms of interaction forces and potential energies. 
The correct quantitative form of this interaction, however, has remained an open question.
Here, we introduce a novel statistical-mechanical approach to directly measure the interaction energy between pedestrians.  
This analysis, when applied to a large collection of human motion data, reveals a simple power law interaction that is based not on the physical separation between pedestrians but on their projected time to a potential future collision, and is therefore fundamentally anticipatory in nature. 
Remarkably, this simple law is able to describe human interactions across a wide variety of situations, speeds and densities. We further show, through simulations, that the interaction law we identify is sufficient to reproduce many known crowd phenomena.  
\end{abstract}


\maketitle

In terms of its large-scale behaviors, a crowd of pedestrians can look strikingly similar to many other collections of repulsively-interacting particles~\cite{HMF01,MGM+12,YJ07,Helb01}. 
These similarities have inspired a variety of pedestrian crowd models, including cellular automata and continuum-based approaches \cite{K85,HB00,HUG02,KNS03}, as well as simple particle or agent-based models \cite{HFV00,YCD+05,CSS10,HB03,MHJ+09,fajen2003behavioral,boids}.
Many of these models conform to a long-standing hypothesis that humans in a crowd interact with their neighbors through some form of ``social potential'' \cite{HM95}, analogous to the repulsive potential energies between physical particles. How to best determine the quantitative form of this interaction potential, however, has remained an open question, with most previous researchers employing a simulation-driven approach.

\begin{figure}
\includegraphics[]{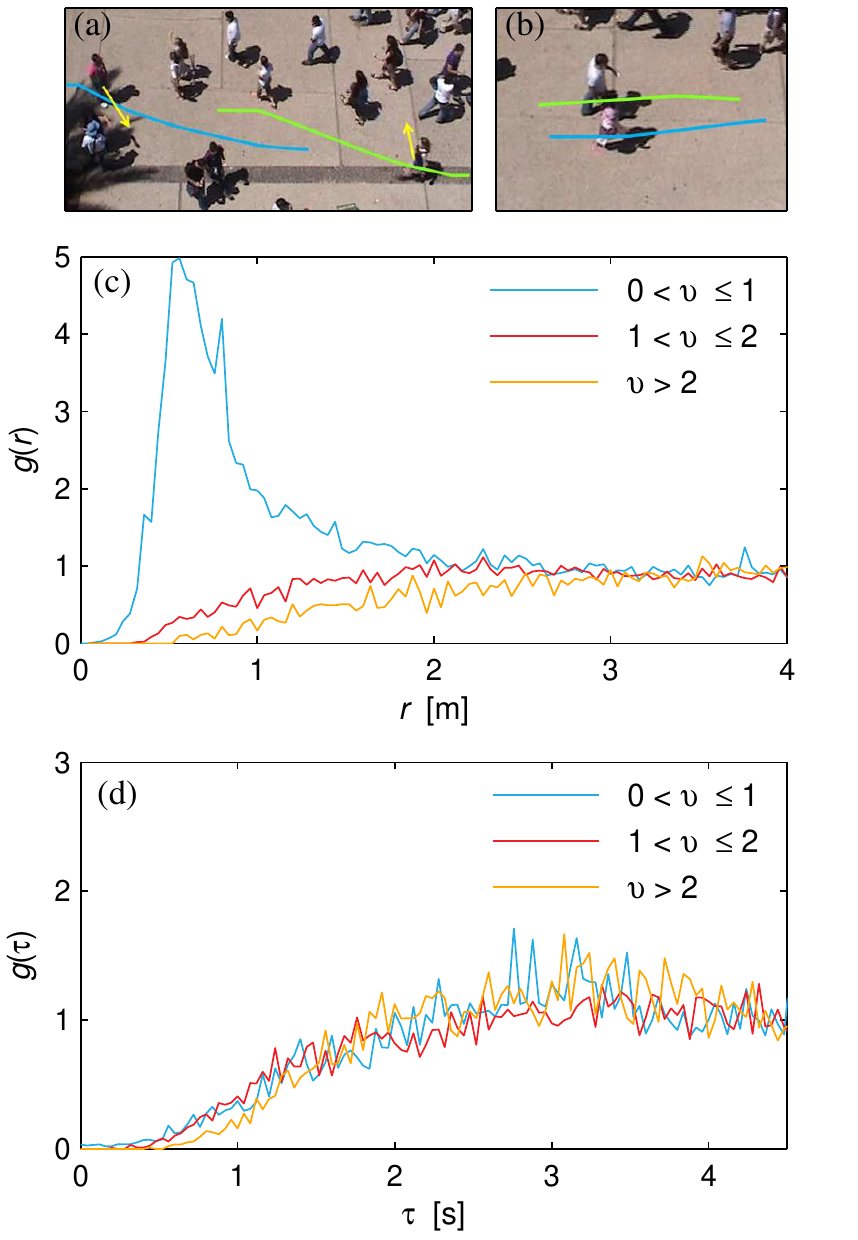}
\caption{
Analysis of anticipation effects in pedestrian motion.
(a) Two pedestrians react strongly to avoid an upcoming collision even though they are far from each other (path segments over an interval of $4\,$s 
are shown as colored lines, with arrows indicating acceleration). (b) In the same environment, two pedestrians walk close to each other without any relative acceleration. (c) The pair distribution function $g$ as a function of inter-pedestrian separation $r$ shows very different behavior when plotted for pedestrian pairs with different rate of approach $v = -dr/dt$. Units of $v$ are m/s. 
(d) In contrast, when $g$ is computed as a function of time-to-collision, $\tau$, curves corresponding to different $v$ collapse onto each other.
}
\label{fig:fig1}
\end{figure}

Previously, direct measurement of the interaction law between pedestrians has been confounded by two primary factors. First, each individual in a crowd experiences a complex environment of competing forces, making it difficult to isolate and robustly quantify a single pairwise interaction. Secondly, a pedestrian's motion is strongly influenced not just by the present position of neighboring pedestrians, but by their anticipated future positions \cite{GLRM05,DVKB05,Joh09,NDA11,OMCP12}, a fact which has influenced recent models \cite{OPOD10,zanlungo2011social,MHT11,GCL+12}. Consider, for example, two well-separated pedestrians walking into a head-on collision (\figref{fig:fig1}a). These pedestrians typically exhibit relatively large acceleration as they move to avoid each other, as would result from a large repulsive force. On the other hand, pedestrians walking in parallel directions exhibit almost no acceleration, even when their mutual separation is small (\figref{fig:fig1}b). 

Here, we address both of the aforementioned factors using a data-driven, statistical mechanics-based analysis that accounts properly for the anticipatory nature of human interactions.  
This approach allows us to directly and robustly measure the interaction energy between pedestrians. 
The consistency of our measurements across a variety of settings suggests a simple and universal law governing pedestrian motion. 

To perform our analysis, we turn to the large collections of recently published crowd datasets recorded by motion-capture or computer vision-based techniques. These datasets include pedestrian trajectories from several outdoor environments \cite{LCL07,PES09} and controlled lab settings \cite{SPS+09} (a summary of datasets is given in 
the Supplemental Material \cite{sup}). To reduce statistical noise, datasets with similar densities were combined together, resulting in one \emph{Outdoor} dataset comprising 1,146 trajectories of pedestrians in sparse-to-moderate outdoor settings, and one \emph{Bottleneck} dataset with 354 trajectories of pedestrians in dense crowds passing through narrow bottlenecks. In analyzing these datasets, our primary tool for quantifying the strength of interactions between pedestrians is the statistical-mechanical pair distribution function, denoted~$g$.

As in the typical condensed matter setting \cite{mahan1990}, here we define the pair distribution function $g(x)$ as the observed probability density for two pedestrians to have relative separation $x$ divided by the expected probability density for two non-interacting pedestrians to have the same separation.
In general, the probability density for non-interacting pedestrians cannot be known \emph{a priori}, since it depends on where and how frequently pedestrians enter and exit the environment. However, for large datasets we are able to closely approximate this distribution by sampling the separation between all pairs of pedestrians that are not simultaneously present in the scene (and therefore not interacting). As defined above, small values of the pair distribution function, $g(x)\ll 1$, correspond to situations where interactions produce strong avoidance.

\IGNORE{
If the Cartesian distance $r$ between two pedestrians were a sufficient descriptor of their interaction, we would expect the shape of the pair distribution function $g(r)$ to be independent of all other variables, such as the rate at which the two pedestrians are approaching each other, $v = -dr/dt$.
However, as discussed above, the distance $r$ between pedestrians cannot by itself accurately parameterize pedestrian interactions. This can be clearly seen in \figref{fig:fig1}c, which shows that $g(r)$ has large, qualitative differences when the data is binned by the value of $v$. 
In particular, pedestrians with a small rate of approach are more likely to be found close together than those that are approaching each other quickly (as evidenced by the separation between the curves at small $r$).
The peak in the curve corresponding to small $v$ suggests a tendency for pedestrians with similar velocity to walk together.
}

If the Cartesian distance $r$ between two pedestrians were a sufficient descriptor of their interaction, we would expect the shape of the pair distribution function $g(r)$ to be independent of all other variables. 
However, as can be seen in \figref{fig:fig1}c, $g(r)$ has large, qualitative differences when the data is binned by the rate at which the two pedestrians are approaching each other, $v = -dr/dt$. 
In particular, pedestrians with a small rate of approach are more likely to be found close together than those that are approaching each other quickly (as evidenced by the separation between the curves at small $r$). A particularly pronounced difference can be seen for the curve corresponding to small $v$, where the large peak suggests a tendency for pedestrians with similar velocities to walk closely together.

While the distance $r$ is not a sufficient descriptor of interactions, we find that the pair distribution function can, in fact, be accurately parameterized by a single variable that describes how imminent potentially upcoming collisions are. 
We refer to this variable as the \emph{time-to-collision}, denoted $\tau$, which we define as the duration of time for which two pedestrians could continue walking at their current velocities before colliding. 
As shown in \figref{fig:fig1}d, when the pair distribution function is plotted as a function of $\tau$, curves for different rates of approach collapse onto each other, with no evidence of a separate dependence of the interaction on $v$. 
Even when binned by other parameters such as the relative orientation between pedestrians, there is no significant difference between curves (see the Supplemental Material~\cite{sup}).
This consistent collapse of the curves suggests that the single variable $\tau$ provides an appropriate description of the interaction between pedestrians.

This pair distribution function, $g(\tau)$, describes the extent to which different configurations of pedestrians are made unlikely by the mutual interaction between pedestrians. In general, situations with strong interactions (small $\tau$) are suppressed statistically, since the mutual repulsion between two approaching pedestrians makes it very unlikely that the pedestrians will arrive at a situation where a collision is imminent.  This suppression can be described in terms of a pedestrian ``interaction energy" $E(\tau)$.  In particular, in situations where the average density of pedestrians does not vary strongly with time, 
the probability of a pair of pedestrians having time-to-collision $\tau$ can be assumed to follow a Boltzmann-like relation, $g(\tau) \propto \exp[-E(\tau)/E_0]$.  Here, $E_0$ is a characteristic pedestrian energy, whose value is scene-dependent.  

This use of a Boltzmann-like relation between $g(\tau)$ and $E(\tau)$ amounts to an assumption that the systems being considered are at, or near, statistical equilibrium. 
In our analysis, this assumption is motivated by the observation that the intensive properties of the system in each of the datasets (e.g., the average pedestrian density and walking speed) are essentially time-independent. If this time-independence is taken as given, a Boltzmann-like relation follows as a consequence of entropy maximization. 
By rearranging this relation, the interaction energy can be expressed in terms of $g(\tau)$ as: 
\begin{equation}
E(\tau) \propto \ln{[1/g(\tau)]}.
\label{eq:eq1}
\end{equation}
A further, self-consistent validation of \eq{eq1} is provided below.

\begin{figure}
\centering
\includegraphics[]{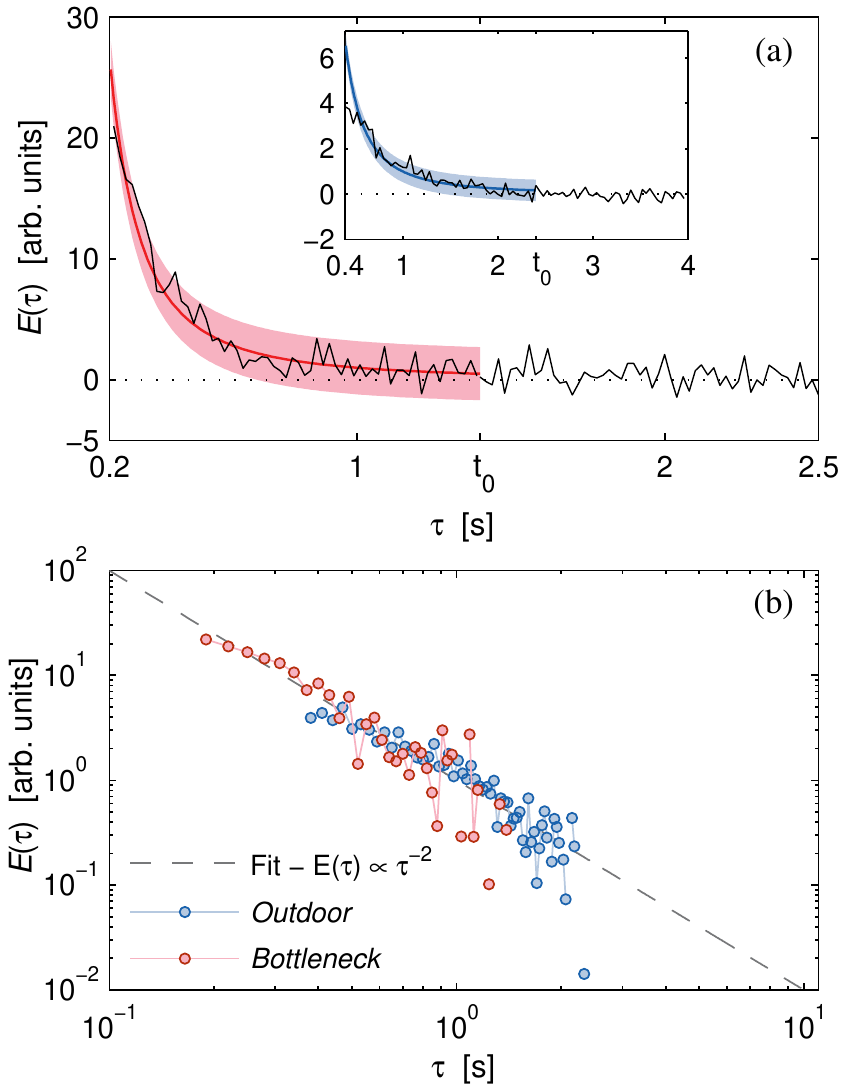}
\caption{
(a) The interaction energy computed from the dense {\it Bottleneck} dataset and from the more sparse {\it Outdoor} dataset (inset).  The overall constant $k$ is normalized so that $E(1)=1$.  Both datasets fit well to a power law up to a point marked $t_0$, beyond which there is no discernible interaction. Solid lines shows the fit to the data and colored regions show their corresponding 95\% confidence interval ({\it Bottleneck}, $R^2 = 0.94$; {\it Outdoor}, $R^2 = 0.92$).  (b) The interaction energy in  both datasets is well described by a power law with exponent $2$. 
\label{fig:fig2}
}
\end{figure}

Figure~\ref{fig:fig2} plots the interaction law defined by \eq{eq1} using the values for $g(\tau)$ derived from our two aggregated pedestrian datasets. It is worth emphasizing that these two datasets capture very different types of pedestrian motion. 
The pedestrian trajectories in the \emph{Outdoor} dataset are generally multi-directional paths in sparse-to-moderate densities, with pedestrians often walking in groups or stopping for brief conversations. In contrast, trajectories in the \emph{Bottleneck} dataset are largely unidirectional, with uniformly high density, and with little stopping or grouping between individuals.

Remarkably, despite their large qualitative differences, both datasets reveal the same power-law relationship underlying pedestrian interactions. For both datasets, the interaction energy $E$ shows a quadratic falloff as a function of $\tau$, so that ${E(\tau)\propto 1/\tau^{\scriptstyle2}}$ 
over the interval where $E$ is well-defined. 
For smaller values of $\tau$ (less than $\sim$200 ms), the energy seen in the data saturates to a maximum value, likely as a consequence of finite human reaction times. 
For sufficiently large values of $\tau$, on the other hand, the observed interaction energy quickly vanishes, suggesting a truncation of the interaction when the time-to-collision is large. We denote the maximum observed interaction range as $t_0$ (\emph{Bottleneck: }$t_0\approx 1.4\,\mathrm{s}$; \emph{Outdoor: }$t_0 \approx 2.4\,\mathrm{s}$). 

Importantly, $t_0$ does not, by itself, indicate the intrinsic interaction range between pedestrians, since interactions between distant, non-neighboring pedestrians are screened by the presence of nearest-neighbors, as in other dense, interacting systems \cite{mahan1990, BSF+13}.  
For a crowd with density $\rho$, the characteristic ``screening time" can be expected to scale 
as the typical distance between nearest neighbors, $\rho^{-1/2}$, divided by the mean walking speed $u$.  Such scaling is indeed consistent with the trend observed in our data, with the denser \emph{Bottleneck} dataset ($\rho = 2.5$\,m$^{-2}$, $u = 0.55$\,m/s) demonstrating a smaller value of $t_0$ than the sparser \emph{Outdoor} dataset ($\rho = 0.27$\,m$^{-2}$, $u = 0.86$\,m/s)
\footnote{The ratio of $(\rho^{-1/2})/u$ between the \emph{Bottleneck} and \emph{Outdoor} datasets is $(1.15\,\mathrm{s})/(2.25\,\mathrm{s}) = 0.51$. For comparison, the measured values of $t_0$ have a ratio $(1.4\,\mathrm{s})/(2.4\,\mathrm{s}) = 0.58$.}.   
While the large-$\tau$ behavior in our datasets is therefore dominated by screening, we can use the largest observed values of $t_0$ to place a lower bound estimate on the intrinsic range of unscreened interactions (which we denote as $\tau_0$).  This estimate suggests that an appropriate value is $\tau_0 \approx 3$\,s, which is consistent with previous research demonstrating an interaction time horizon of $2 - 4\,\mathrm{s}$ \cite{OMCP12}. 

\IGNORE{
Importantly, $t_0$ does not, by itself, indicate the \emph{intrinsic} interaction range between pedestrians, since interactions between distant, non-neighboring pedestrians are screened by the presence of nearest-neighbors, as in other densely-interacting systems \cite{mahan1990, BSF+13}.  
Typically, one can expect the characteristic ``screening time" in a dense system to scale as $t_0 \sim n^{-1/2}/s$, where $n$ is the pedestrian density and $s$ is the \dcomment{average} walking speed, so that more dense environments produce stronger screening.  This scaling is consistent with the trend observed in our data, with the denser \emph{Bottleneck} dataset \dcomment{($n = 2.5$\,m$^{-2}$, $s = 0.55$\,m/s)} demonstrating a smaller value of $t_0$ than the sparser \emph{Outdoor} dataset \dcomment{($n = 0.27$\,m$^{-2}$, $s = 0.86$\,m/s)}.  
Consequently, the intrinsic range of unscreened pedestrian interactions (which we denote as $\tau_0$), can be bounded from below by the largest observed values of $t_0$, suggesting that an appropriate value is $\tau_0\approx 3\,\mathrm{s}$. This is consistent with previous research demonstrating an interaction time horizon of $2 - 4\,\mathrm{s}$ \cite{OMCP12}.
}

Since the interaction energy follows a power law with a sharp truncation at large $\tau$, we infer from the data the following form of the pedestrian interaction law:
\begin{equation}
E(\tau)=\frac{k}{\tau^2}e^{-\tau/\tau_0}.
\label{eq:eq2}
\end{equation}
Here, $k$ is a constant that sets the units for energy.

To demonstrate the general nature of the identified interaction law, we performed simulations of pedestrians that adapt their behavior according to \eq{eq2} via force-based interactions. In particular, the energy $E(\tau)$ directly implies a natural definition of the force $\mathbf{F}$ experienced by pedestrians when interacting:
\begin{equation}
\mathbf{F} = -\nabla_{\mathrm{\mathbf{r}}} \left( \frac{k}{\tau^2} e^{-\tau/\tau_0} \right),
\label{eq:eq3}
\end{equation}
where $\nabla_{\mathrm{\mathbf{r}}}$ is the spatial gradient.
A full analytical expression for this derivative is given in the Supplemental Material~\cite{sup}. 

\begin{figure}
\centering
\includegraphics[]{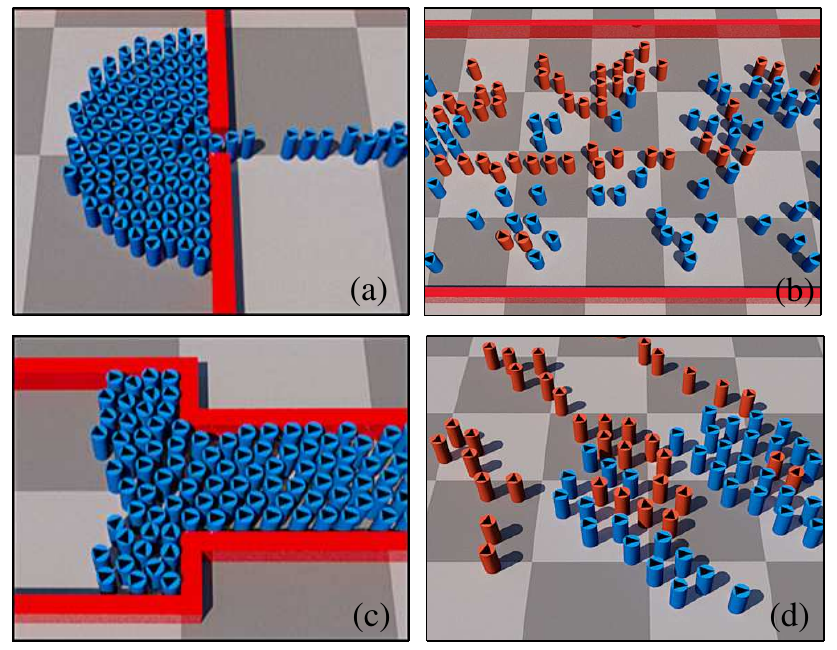}
\includegraphics[]{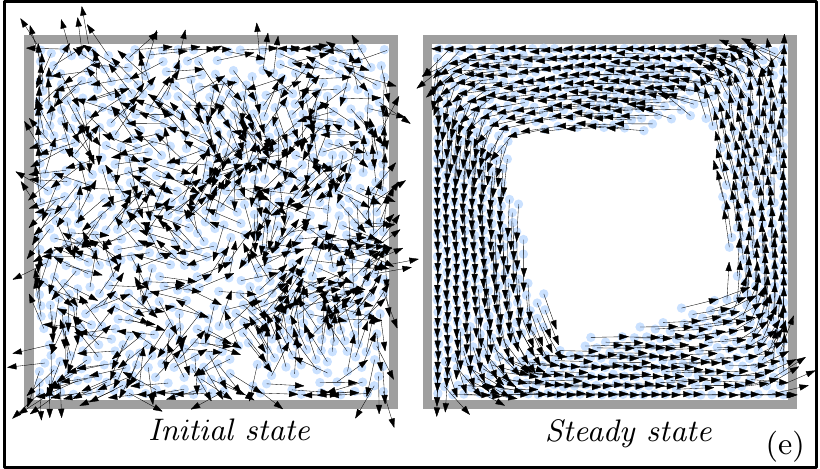}
\caption{
Stills from simulations of agents following the force law derived from \eq{eq2}.  In figures (a)-(d), agents are represented as cylinders and color-coded according to their goal direction. The simulated agents display many emergent phenomena also seen in human crowds, including  arching around narrow passages (a), clogging and ``zipping" patterns at bottlenecks (c), and spontaneously self-organized lane formation (b and d).  Figure (e) depicts a simulation of agents without a preferred goal direction (arrows represent the agents' current orientations). The agents' interactions lead to large-scale synchronization of their motion. 
Further simulation details are given in the Supplemental Material \cite{sup}.
}
\label{fig:fig3}
\end{figure}

\begin{figure}[t]
\centering
\includegraphics[]{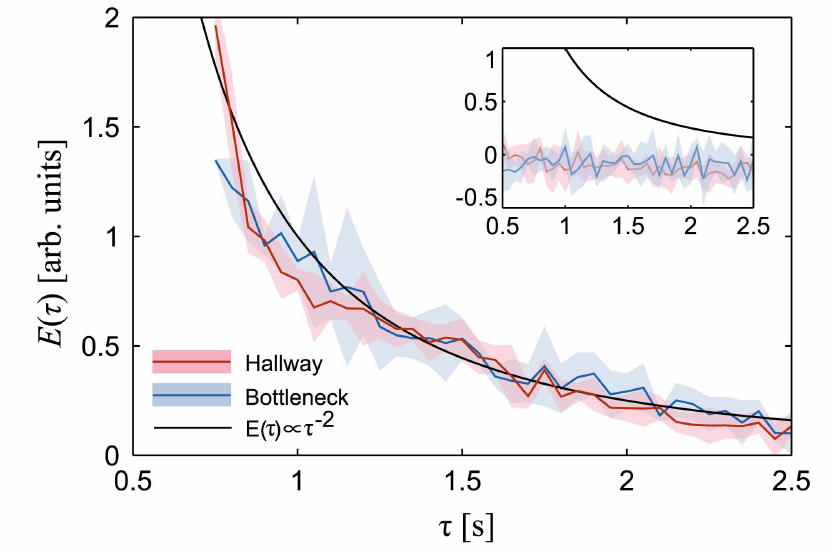}
\caption{
Inferred interaction energy $E \propto \ln(1/g)$ as a function of time-to-collision $\tau$ for different simulations, obtained using 
the anticipatory force described by \eq{eq3}, and  the distance-dependent force described in Ref.\ \onlinecite{HFV00} (inset). 
For simulations with strictly distance-dependent interactions, the inferred interaction energy does not show a dependence on $\tau$. In contrast, simulations following our model closely match the observed empirical power law for $E(\tau)$. 
Shaded regions denote average energy values $\pm$ one standard deviation. 
}
\label{fig:fig4}
\end{figure}

\IGNORE{
For the purposes of simulation, each pedestrian is also given a driving force associated with its desired direction of motion \cite{HFV00}. 
The resulting force model is sufficient to reproduce a wide variety of important pedestrian behaviors, including the formation of lanes, arching in narrow passages, slowdowns in congestion, and anticipatory collision avoidance (\figref{fig:fig3}). 
Additionally, the simulated pedestrians match the empirically known \textit{fundamental diagram} \cite{weidmann} of speed-density relationships for real human crowds.   
Our simulations also reproduce the anticipatory power law described by \eq{eq2}, as shown in \figref{fig:fig4}, when used to model situations that are very similar to those from which the real data was recorded.  This provides a self-consistent validation of our use of the Boltzmann relation, \eq{eq1}, to infer the interaction energy from data.  Our simulations also generally reproduce the empirical behavior of $g(r)$ depicted in \figref{fig:fig1}c \cite{sup}.
In contrast, simulations 
generated by distance-based interaction forces 
fail to show a dependence of $E$ on $\tau$ (\figref{fig:fig4}b), or to accurately reproduce $g(r)$. Other more recent models of pedestrian behavior also cannot consistently capture the power-law relationship observed in the data \cite{sup}. 
}

For the purposes of simulation, each pedestrian is also given a driving force associated with its desired direction of motion, following Ref.~\onlinecite{HFV00}. 
The resulting force model is sufficient to reproduce a wide variety of important pedestrian behaviors, including the formation of lanes, arching in narrow passages, slowdowns in congestion, and anticipatory collision avoidance (\figref{fig:fig3}). Additionally, the simulated pedestrians match the known \textit{fundamental diagram} \cite{weidmann} of speed-density relationships for real human crowds and qualitatively capture the empirical behavior of $g(r)$ depicted in \figref{fig:fig1}c \cite{sup}. 

Our simulations also reproduce the anticipatory power law described by \eq{eq2}, as shown in \figref{fig:fig4}.  
In contrast, simulations generated by distance-based interaction forces fail to show a dependence of $E$ on $\tau$ (\figref{fig:fig4}). Other, more recent models of pedestrian behavior also cannot consistently capture the empirical power-law relationship (see the Supplemental Material~\cite{sup}).  The ability of our own simulations to reproduce $E(\tau)$ also provides a self-consistent validation of our use of the Boltzmann relation to infer the interaction energy from data.

Interesting behavior can also be seen when \eq{eq3} is applied to 
walkers propelled forward in the direction of their current velocity without having a specific goal  
(as implemented, for example, in Ref.\ \onlinecite{GAJ08}). 
In such cases, complex spatio-temporal patterns 
emerge, leading eventually to large scale synchronization of motion. 
An example is illustrated in \figref{fig:fig3}e, where a collection of pedestrians that is initialized to a high energy state with many imminent collisions  settles over time into a low energy state where pedestrians move in unison.  
This result is qualitatively similar to observed behavior in dense, non-goal-oriented human crowds \cite{SBS+13}, and is reminiscent of the ``flocking" behavior seen in a variety of animal groups \cite{VCB+95, CKJ+02, BCC+08, BH09, VZ12}.  A detailed study of such collective behaviors, however, is outside the scope of our present work.

While the model implied by \eq{eq3} is widely applicable, it may not be sufficient on its own to capture certain crowd phenomena, such as the shock waves and turbulent flows that have been reported to occur in extremely high density crowds \cite{HJA07}. 
In such very dense situations, saturating effects such as finite human reaction time become relevant, and these may alter the quantitative form of the interaction in a way that is not well-captured by our time-to-collision-based analysis.  Augmenting our result with an additional close-ranged component of the interaction may give a better description of these extremely dense scenarios, and is a promising avenue for future work.

To conclude, our statistical mechanics-based analysis of a large collection of human data has allowed us to quantify the nature and strength of interactions between pedestrians. 
This novel type of analysis opens new avenues for studying the behavior of humans using real life data. 
The data we have analyzed here reveals the existence of a single anticipatory power law governing the motions of humans. 
The consistency of this law across a variety of scenarios provides a new means to understand how pedestrians behave and suggests new ways to evaluate models of pedestrian interactions. 
Further, these results suggest a general quantitative law for describing human anticipation that may extend to other studies of human behavior, which may therefore be amenable to a similar type of analysis. 

\acknowledgments
\vspace{1ex}
Complete simulation source code, along with videos and links to data used in this study, can be found at our companion webpage: \url{http://motion.cs.umn.edu/PowerLaw}. 
We would like to thank Anne-H\'{e}l\`{e}ne Olivier, Alex Kamenev, Julien Pettr\'{e}, Igor Aranson, Dinesh Manocha, and Leo Kadanoff for helpful discussions.  We also acknowledge support from Intel and from University of Minnesota's MnDRIVE Initiative on Robotics, Sensors, and Advanced Manufacturing.  Work at Argonne National Laboratory is supported by the U.S. Department of Energy, Office of Basic Energy Sciences under contract no. DE-AC02-06CH11357.

\bibliographystyle{apsrev4-1}
\bibliography{bibliography}

\end{document}


\title{Supplemental material for: \\ Universal power law governing pedestrian interactions}

\author{Ioannis Karamouzas}
\affiliation{Department of Computer Science and Engineering, University of Minnesota, USA}
\author{Brian Skinner}
\affiliation{Materials Science Division, Argonne National Laboratory, USA}
\author{Stephen J. Guy}
\affiliation{Department of Computer Science and Engineering, University of Minnesota, USA}

\date{\today}

\maketitle

\onecolumngrid 

\section{Experimental datasets}
\begin{figure*}[b]
\centering
\includegraphics[width=0.9\textwidth]{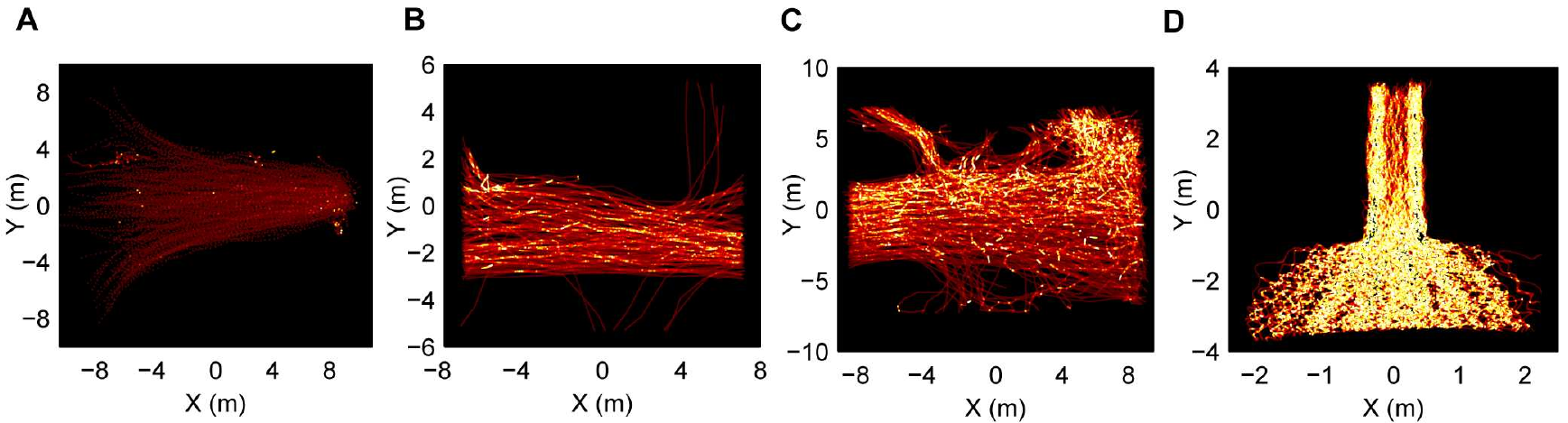}
\caption{
Pedestrian trajectories for four of the six datasets examined in this paper, colored by time-averaged density from low (dark red) to high (white) values. (A-B) are from sparse, outdoor environments with largely bi-directional flows. (C) is from a moderately dense environment with multi-directional flow. (D) is from controlled experiments where a dense crowd walks through a narrow constriction. 
Horizontal and vertical axes label the distance from the center of the scene, in meters.
}
\label{fig:figS1}
\end{figure*}

Our findings draw from datasets published by three different research groups.  These datasets are illustrated in \figref{fig:figS1} and summarized in Table~\ref{tab:tabS1}. The experiments labeled b250\_combined and b4\_combined are described in Refs.~\onlinecite{SPS+09, Seyfried08}, and are combined to comprise the \emph{Bottleneck} dataset. These experimental trials were recorded at 25\,fps using multiple cameras, with trajectories automatically tracked and corrected for perspective distortion \cite{SPS+09,BSR+10}. 
The experiments involved participants walking through a 4$\,$m-long corridor that has a bottleneck of width of 2.5$\,$m or 1$\,$m. 
The remaining four datasets were combined to comprise the {\it Outdoor} dataset. The three datasets labeled crowds\_zara01, crowds\_zara02, and students003 were recorded at 25\,fps using a single camera and the trajectories of the pedestrians
were manually tracked and post-processed to minimize errors and correct the distortion due to pixel noise and the camera perspective \cite{LCL07}. 
The dataset denoted seq\_eth was recorded at 2.5\,fps and tracked in a semi-automatic process \cite{PES09}.
In addition to their different settings, the two datasets types, {\it Bottleneck} and {\it Outdoor}, are also distinguished from each other by their different pedestrian densities (see Table \ref{tab:tabS1}), while within each type the densities are similar.  

\begin{table*}
\caption{
Characteristics of the datasets analyzed. The first two listed datasets were grouped together to comprise the {\it Bottleneck} dataset. The remaining four datasets define the combined {\it Outdoor} dataset. The average density reported in the table is computed using the generalized definition of Edie \cite{Edie63}. 
}
\

\renewcommand{\arraystretch}{.5}
\setlength{\tabcolsep}{6.0pt}
\centering
\begin{tabular}{|c|c|c|c|c|c|c|c|c|l|}
\hline
\multirow{2}{*}{\textbf{Dataset}} & \multirow{2}{*}{\textbf{Type}} & \multirow{2}{*}{\textbf{Description}} & \multirow{2}{*}{\textbf{Flow}}
& \multirow{2}{*}{\textbf{Location}} &  \textbf{No.} &  \textbf{Density} &  \multirow{2}{*}{\textbf{Citation}} & \textbf{Data}\\
& & & & & \textbf{Ped} & $\bf (1/\mathrm{\textbf{m}}^2)$ & & \textbf{files}\\
\hline
 &  & \multicolumn{1}{l|}{$\ \ $Participants} & & & \multirow{4}{*}{176}& \multirow{4}{*}{2.328}&\multirow{4}{*}{\cite{SPS+09,Seyfried08}} &\multirow{4}{*}{\cite{web1}} \\
b250\_& Lab &  \multicolumn{1}{l|}{$\ \ $navigating} & Uni- &J{\"u}lich, & & & &\\
combined& setting&  \multicolumn{1}{l|}{$\ \ $through a 2.5$\,$m} & directional & Germany& & & &\\
& &  \multicolumn{1}{l|}{$\ \ $wide bottleneck} & & & & & &\\
\hline

&  & \multicolumn{1}{l|}{$\ \ $Participants} & & & \multirow{4}{*}{178}& \multirow{4}{*}{2.665}&\multirow{4}{*}{\cite{SPS+09,Seyfried08}} &\multirow{4}{*}{\cite{web1}} \\
l4\_& Lab &  \multicolumn{1}{l|}{$\ \ $navigating} & Uni- &J{\"u}lich, & & & &\\
combined& setting&  \multicolumn{1}{l|}{$\ \ $through a 1.5$\,$m} & directional & Germany& & & &\\
& &  \multicolumn{1}{l|}{$\ \ $wide bottleneck} & & & & & &\\
\hline

\multirow{2}{*}{crowds\_}& \multirow{2}{*}{Outdoor} & \multicolumn{1}{l|}{$\ \ $Pedestrians} &\multirow{2}{*}{Bi-} &\multirow{2}{*}{Nicosia,} & \multirow{3}{*}{148}& \multirow{3}{*}{0.206}&\multirow{3}{*}{\cite{LCL07}} &\multirow{3}{*}{\cite{web2}} \\
\multirow{2}{*}{zara01}& \multirow{2}{*}{setting} &  \multicolumn{1}{l|}{$\ \ $interacting at a} &\multirow{2}{*}{directional} &\multirow{2}{*}{Cyprus} & & & &\\
& & \multicolumn{1}{l|}{$\ \ $shopping street} & & & & & &\\
\hline

\multirow{2}{*}{crowds\_}& \multirow{2}{*}{Outdoor} & \multicolumn{1}{l|}{$\ \ $Pedestrians} &\multirow{2}{*}{Bi-} &\multirow{2}{*}{Nicosia,} & \multirow{3}{*}{204}& \multirow{3}{*}{0.267}&\multirow{3}{*}{\cite{LCL07}} &\multirow{3}{*}{\cite{web2}} \\
\multirow{2}{*}{zara02}& \multirow{2}{*}{setting} &  \multicolumn{1}{l|}{$\ \ $interacting at a} &\multirow{2}{*}{directional} &\multirow{2}{*}{Cyprus} & & & &\\
& & \multicolumn{1}{l|}{$\ \ $shopping street} & & & & & &\\
\hline

\multirow{3}{*}{students003}& \multirow{2}{*}{Outdoor} & \multicolumn{1}{l|}{$\ \ $Students} &\multirow{2}{*}{Multi-} &\multirow{2}{*}{Tel Aviv,} & \multirow{3}{*}{434}& \multirow{3}{*}{0.391}&\multirow{3}{*}{\cite{LCL07}} &\multirow{3}{*}{\cite{web2}} \\
& \multirow{2}{*}{setting} &  \multicolumn{1}{l|}{$\ \ $interacting at a} &\multirow{2}{*}{directional} &\multirow{2}{*}{Israel} & & & &\\
& & \multicolumn{1}{l|}{$\ \ $university campus} & & & & & &\\
\hline

\multirow{4}{*}{seq\_eth}&  & \multicolumn{1}{l|}{$\ \ $Students} & & & \multirow{4}{*}{360}& \multirow{4}{*}{0.148}&\multirow{4}{*}{\cite{PES09}} &\multirow{4}{*}{\cite{web3}} \\
& Outdoor &  \multicolumn{1}{l|}{$\ \ $interacting outside} &Bi- &Z{\"u}rich, & & & &\\
& setting & \multicolumn{1}{l|}{$\ \ $the ETH  main} &directional &Switzerland & & & &\\
& & \multicolumn{1}{l|}{$\ \ $building} & & & & & &\\
\hline
\end{tabular}
\label{tab:tabS1}
\end{table*}

Matlab was used to process the corresponding 2D positional data of the pedestrians' trajectories after applying a low-pass second order Butterworth filter to remove noise and reduce oscillation effects (zero phase shift; 0.8 normalized cutoff frequency for the four outdoor datasets and 0.24 normalized cutoff frequency for the two bottleneck datasets). 
In all datasets and for each time instant, we infer the instantaneous velocity of each pedestrian using a discrete derivative. 
To estimate the time-to-collision $\tau$, which indicates when (if ever) a given pair of pedestrians will collide if they continue moving with their current velocities, we assume that each pedestrian can be modeled as a disc with a fixed radius. In our analysis, we used a radius of $0.1$\,m, which results in less than ten total colliding pairs of pedestrians across all six datasets. The corresponding disc diameter represents the smallest cross-sectional length that a pedestrian can occupy (by rotating or compressing his or her upper body) while resolving a collision. 
Overall, we collected 119,774 pairwise time-to-collision ({$\tau$}) samples from the {\it Outdoor} datasets and 177,672 samples from the {\it Bottleneck} datasets. 
This number of samples was sufficient to draw statistically significant conclusions about the power law governing pedestrian interactions. While all results presented here correspond to an assumed radius of $0.1$\,m, a similar power law trend is produced for a wide range of radius values.

\section{Detailed description of the pair-distribution function}
For a given variable \mm{$x$} that describes the separation between two pedestrians, the pair distribution function {$g(x)$} indicates the degree to which a pair separation $x$ is made unlikely by the interactions between pedestrians. Specifically, {$g(x)$} is defined by $g(x) =$ \mm{$P(x)/P_\text{NI}(x)$}, where \mm{$P(x)$} is the probability density function for the relative separation {$x$} between pedestrians in the dataset, and \mm{$P_\text{NI}(x)$} is the probability density function for $x$ that would arise, hypothetically, if pedestrians were non-interacting. For large separation values we expect that pedestrians do not influence each other, and therefore  \mm{$\underset{x\to\infty}{\lim}g(x)=1$}.

While there is no way to remove the interactions between pedestrians in the dataset, we propose the following approach to closely approximate the distribution \mm{$P_\text{NI}(x)$}. 
We begin by randomly permuting time information between different pedestrians, so that each moment in a pedestrian's trajectory has its instantaneous position and velocity preserved, but is assigned a randomly permuted time. The resulting ``time-scrambled'' dataset maintains the same spatially-averaged density at any given instant and the same time-averaged flow rate of pedestrians across any location in the scene as in the original dataset. However, the pedestrian positions in the time-scrambled dataset are uncorrelated with each other at any given value of the {\it scrambled} time, since they are drawn from trajectories at different {\it real} times. Creating a probability density function of the scrambled data gives \mm{$P_\text{NI}(x)$}, allowing us to define $g(x) =$ \mm{$P(x)/P_\text{NI}(x)$}. 
Figure~1c of the main text shows the result of this process using for the variable {$x$} the Cartesian distance $r$ between pedestrians, while Fig.~1d shows the result when the probability density functions are computed for the time-to-collision {$\tau$}. 

\section{Statistical methods}
\subsection{Analysis of similarity between $g(r)$ and $g(\tau)$ curves} 
We analyzed the effect that the relative velocity between pedestrians has on the distance- and time-to-collision-based pair distribution functions by conducting separate one-way ANOVAs for $g(r)$ and $g(\tau)$ respectively on the {\it Outdoor} dataset. All tests were performed in STATISTICA (version 8.0) with significance levels set to 5\%. To estimate $g(r)$ and $g(\tau)$, we used intervals of 0.04$\,$m and 0.04$\,$s, respectively, and clustered pairs of pedestrians into three categories according to their rate of approach $v = -dr/dt$ (Fig.~1c and Fig.~1d) for values of $r<8\,$m and $\tau<8\,$s. The analysis reveals that the distance-based pair distribution function $g(r)$ has a significant dependence on the rate of approach, $\left[F(2, 576) = 27.811, P < 0.001\right]$.
However, for $g(\tau)$, the pair distribution function does not vary for different values of $v$, $\left[F(2, 510) = 0.143, P = 0.866\right]$, indicating that $\tau$ is a sufficient descriptor of pedestrian interactions. 

\subsection{Power-law fit for $E(\tau)$} 
Regarding the power-law fit shown in Fig.~2 of the main text, for both the {\it Outdoor} and the {\it Bottleneck} datasets, we estimated $g(\tau)$, and subsequently the interaction energy $E(\tau)$, using intervals of 0.01$\,$s. In both datasets, due to tracking errors and statistical noise, the energy is only well-defined over a finite interval, with the computed energy values fluctuating around a maximum observed energy at very small values of $\tau$ and becoming indistinguishable from noise at large values of $\tau$. To estimate the lower $\tau$ boundary, we first clustered the data into bins of 0.2$\,$s, and used a series of t-tests between successive bins to determine the first two bins with significantly different interaction energies. In the {\it Outdoor} dataset, the analysis revealed a significant difference in $E(\tau)$
between [0.2$\,$s, 0.4$\,$s) and [0.4$\,$s, 0.6$\,$s) indicating a value of $\tau=0.4\,$s as an appropriate lower bound $\left[t(38) = 6.664, P < 0.001\right]$. In the {\it Bottleneck} dataset, the first two bins already exhibit a 
statistically significant difference in energy $\left[t(38) = 10.856, P < 0.001\right]$, allowing us to select the value of 0.2$\,$s as a lower boundary for $\tau$. 
To estimate the upper boundary value, $t_0$, we conducted separate ANOVA tests and determined the first three successive clusters for which the interaction energy does not vary. In the {\it Outdoor} dataset, [2.2$\,$s, 2.4$\,$s) denotes the first bin that has the same energy as its two subsequent ones, indicating $t_0 = 2.4\,$s $\left[F(2, 57) = 1.883, P = 0.161\right]$. In the {\it Bottleneck} dataset, the corresponding bin is [1.2$\,$s, 1.4$\,$s) resulting in the estimate $t_0 = 1.4\,$s $\left[F(2, 57) = 1.614, P = 0.208\right]$.

Over the interval of well-defined data, $E(\tau)$ follows a power law. A linear fit of $\log E$ vs.\ $\log \tau$ with bisquare weighting reveals an exponent of $2.05\pm0.123$ for the {\it Outdoor} dataset and $2.017\pm0.192$ for the {\it Bottleneck} dataset. 
As can be seen in Fig.~2b, the interaction energy in both datasets can be well modeled with an exponent of 2 $[t(174)~=~0.809, P = 0.42$ for the {\it Outdoor} and $t(106) = 0.171, P = 0.865$ for the {\it Bottleneck}$]$. We note that, for visual clarity, the data in Fig.~2a and Fig.~2b are down-sampled, showing $E(\tau)$ samples every 0.02$\,$s and 0.03$\,$s respectively. 

\begin{figure*}[b]
\centering
\includegraphics[width=0.4\textwidth]{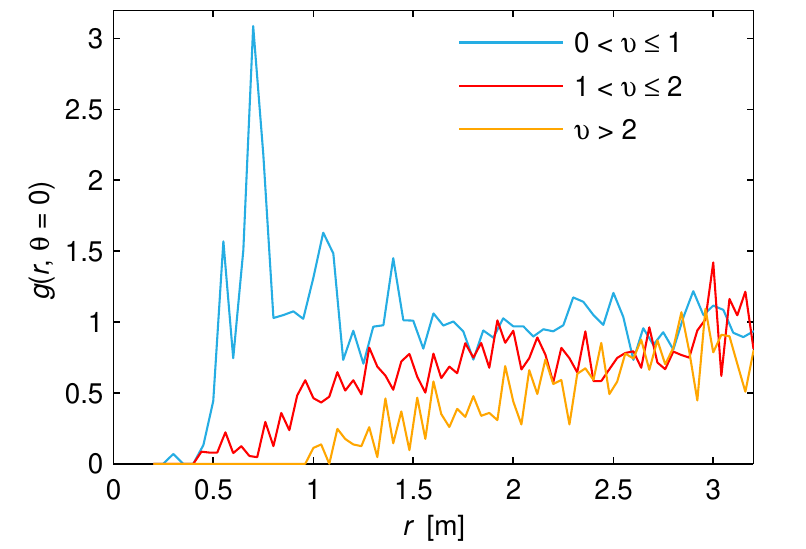}
\caption{
The pair distribution function $g$ in polar coordinates $(r, \theta)$ for pedestrians in the \emph{Outdoor} dataset, plotted for the fixed angle $\theta=0$. $g(r, \theta=0)$ shows a significant dependence on the rate $v$ at which the pedestrians approach each other, indicating that the displacement vector $\rr$ is not a sufficient descriptor of pedestrian interactions.
}
\label{fig:figS2}
\end{figure*}

\section{Pair distribution function in 2D spatial coordinates}
Figure 1c of the main text demonstrates that the Cartesian distance $r$ between two pedestrians is not a sufficient descriptor of their interaction, as it cannot account for the dependence of the pair distribution function $g(r)$ on the pedestrian rate of approach $v$.  Here we show that even the full two-dimensional displacement vector $\rr$ cannot adequately parameterize pedestrian interactions.  In other words, we show that the empirical behavior of the pair distribution function is inconsistent with any form of the interaction that depends only on relative spatial coordinates.

The displacement vector $\rr$ connecting two pedestrians can be parameterized by the vector norm $r$ and the angle $\theta$ between $\rr$ and a given pedestrian's current heading.  
We find that, for a given fixed value of $\theta$, the pair distribution function $g(r, \theta)$ shows a significant dependence on the pedestrian rate of approach $v$.  This is shown explicitly for $\theta = 0$ in \figref{fig:figS2}. 
Similar conclusions can also be drawn for different values of $\theta$, suggesting that the displacement vector $\rr$ alone cannot accurately quantify pedestrian interactions.

\section{Orientation-independence of $g(\tau)$}
In Fig.~1d of the main text, it is shown that $g(\tau)$ is independent of the rate $v$ at which pedestrians approach other. 
Here we show that $g(\tau)$ is also independent of the relative orientation between pedestrians.  

\ffigref{fig:figS3}A plots $g(\tau)$ for different values of the angle $\theta$, defined (as above) as the angle between a pedestrian's velocity vector and the displacement vector connecting the two interacting pedestrians.  As in Fig.~1d, the curves for $g(\tau)$ corresponding to different $\theta$ collapse onto each other.  This suggests that the interaction between pedestrians as captured by $\tau$ is independent of the pedestrians' relative orientation. 

\section{Absence of interaction for undefined $\tau$}
Figure~2 of the main text presents the interaction energy for pedestrian pairs with finite time-to-collision $\tau$.  Here we demonstrate that for pedestrian pairs that are not on a collision course (i.e., for pairs with undefined $\tau$), there is no evidence of any finite interaction beyond a short-ranged exclusion.

\begin{figure*}[t]
\centering
\includegraphics[width=0.85\textwidth]{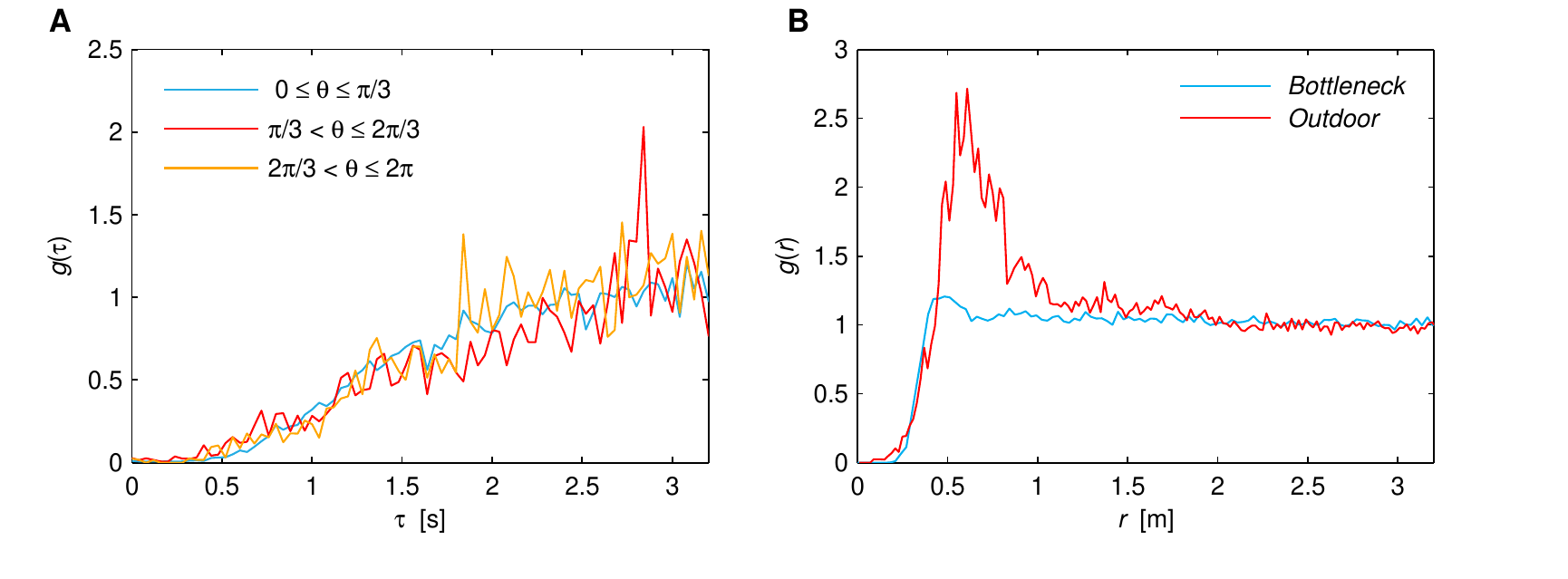}
\caption{
Appropriateness of the time-to-collision variable ($\tau$) as a descriptor of pedestrian-pedestrian interactions. 
(A) The pair distribution function $g(\tau)$ as a function of time-to-collision $\tau$ for different values of the relative orientation $\theta$ between interacting pedestrians in the \emph{Outdoor} dataset. 
The different $g(\tau)$ curves collapse onto each other, suggesting that, in addition to the velocity-independence demonstrated in Fig.\ 1d of the main text, $g(\tau)$ is also independent of the pedestrians' relative orientation. 
(B) The pair distribution function $g$ as a function of the distance $r$ between pedestrian pairs that are not on a collision course (i.e., pairs that have undefined $\tau$). 
For both the \emph{Bottleneck} and the \emph{Outdoor} datasets, there is no evidence of any repulsion beyond $r \approx 0.4$\,m. 
This, along with (B) and Fig.~1 of the main text, suggests that $\tau$ alone provides an appropriate parameterization of pedestrian interactions. 
}
\label{fig:figS3}
\end{figure*}

Figure~\ref{fig:figS3}B shows the pair distribution function $g(r)$ plotted only for pedestrian pairs with undefined $\tau$.  For the {\it Bottleneck} dataset, $g(r) \approx 1$ at all $r \gtrsim 0.4$\,m, which suggests that there is no interaction between non-colliding pedestrians once that are separated by more than $0.4$\,m.  For the {\it Outdoor} dataset, a finite repulsive interaction, corresponding to $g(r) < 1$, also appears only at $r \lesssim 0.4$\,m.  The peak in $g(r)$ at $r \approx 0.6$\,m 
suggests a positive correlation between non-colliding pedestrians that are not too far away. 
This can be mainly attributed to the presence of small groups of pedestrians that walk next to each other, either for social reasons or as a strategy for navigating through the crowd. 

\section{Analytical expression for the simulated interaction force}
Equation (2) of the main text defines the interaction energy of a pair of pedestrians with finite time-to-collision $\tau$.  Within a force-based simulation model, this energy is directly related to the force $\F{ij}$ experienced by the pedestrian {$i$} due to the interaction with another pedestrian {$j$}. In particular, 
\begin{equation}
\F{ij} = -\nabla_{\pos{ij}} E(\tau) = -\nabla_{\pos{ij}}\left( k\tau^{-2}e^{-\tau/\tau_0} \right),
\label{eq:eqS1}
\end{equation}
as in Eq.\ (3) of the main text.
Here, {$\tau$} is understood to be a function of the relative displacement $\pos{ij} = \pos{i} - \pos{j}$ between pedestrians and their relative velocity {$\vel{ij} = \vel{i} - \vel{j}$}. 

At any given simulation step, we estimate {$\tau$} by linearly extrapolating the trajectories of the pedestrians {$i$} and {$j$} based on their current velocities. Specifically, a collision is said to occur at some time {$\tau>0$} if the corresponding discs of the pedestrians, of radii {$R_i$} and {$R_j$}, respectively, intersect. 
If no such time exists, the interaction force $\F{ij}$ is {\bfseries0}. Otherwise, $\tau = \frac{b - \sqrt{d}}{a}$, 
where {$a = \left\|\vel{ij}\right\|^2$}, {$b=-\pos{ij}\cdot\vel{ij}$}, {$c = \left\|\pos{ij}\right\|^2 - (R_i + R_j)^2$}, and {$d = b^2-ac$}.  
By substituting {$\tau$} into \equationref{eq:eqS1}, the interaction force 
can be written explicitly as:
\begin{equation}
\F{ij} = -\left[\frac{ke^{-\tau/\tau_0}}{\left\|\vel{ij}\right\|^2\tau^2}\left(\frac{2}{\tau}+\frac{1}{\tau_0}\right)\right] 
\left[\vel{ij} - \frac{\left\|\vel{ij}\right\|^2\pos{ij}-\left(\pos{ij}\cdot\vel{ij}\right)\vel{ij}}{\sqrt{(\pos{ij}\cdot\vel{ij})^2-\left\|\vel{ij}\right\|^2(\left\|\pos{ij}\right\|^2 - (R_i + R_j)^2)}} \right].
\label{eq:eqS2}
\end{equation}

A complete simulation is produced by combining this interaction force, and a similar force associated with repulsion from static obstacles, with a driving force.  
C++ and Python code of our complete force-based simulation model are available at \url{http://motion.cs.umn.edu/PowerLaw/}.

\section{Simulation results}
We tested the anticipatory interaction law via computer simulations using the derived force-based model. 
To approximate the behavior of typical humans, the preferred walking speeds of the pedestrians were normally distributed with an average value of $1.3 \pm 0.3\,$m/s \cite{weidmann}. In all simulations, we set $k = 1.5$ and $\tau_0 = 3\,$s as the default parameter values of the interaction forces. [See \equationref{eq:eqS2}].

Details of the simulations are listed below:
\begin{itemize}
		\item \emph{Evacuation}: 150 pedestrians exit a room ($10\,$m wide $\times$ $24\,$m long) through a narrow doorway. Due to the restricted movement of the pedestrians, arch-like blockings are formed near the exit, leading to clogging phenomena similar to the ones observed in granular media \cite{HFV00,Helb01}. See Fig.~3a.
	\item \emph{Hallway}: 300 pedestrians cross paths while walking from opposite ends of an open hallway that is $20\,$m wide. The pedestrians dynamically form lanes of uniform walking directions to efficiently resolve collisions. See Fig.~3b.
	\item \emph{Bottleneck}: 150 pedestrians start in a $5\,$m-wide waiting area and have to pass through a $5\,$m-long bottleneck of variable width {($1\,\mathrm{m} -  3\,\mathrm{m}$)}. In all cases, pedestrians exhibit clogging behaviors in the waiting area (Fig.~3c). With wider bottlenecks, ``zipping'' patterns emerge inside the constriction. For example, at a width of $2.5\,$m, pedestrians tend to walk diagonally behind each other, dynamically forming 5-6 overlapping layers that maximize the utility of the bottleneck, as observed in real life \cite{HW05}. 
	\item \emph{Crossing}: Two groups, of 40 pedestrians each, cross paths perpendicularly. The pedestrians prefer to slow down and let others pass rather than deviate from their planned courses. As such, homogeneous clusters of pedestrians emerge within the two groups, leading to the formation of diagonal line-shaped patterns \cite{HD05}. See Fig.~3d. 
	\item \emph{Collective motion}: 750 pedestrians are placed in an enclosed square area of size $40 \times 40$\,m, and at each time step 
are propelled forward in the direction of their current velocity without having a specific goal. 
Pedestrians are initially given random orientations, and after a long enough time they spontaneously form a vortex pattern in which all pedestrians are walking in unison.  See Fig.~3e.
\end{itemize}

\begin{figure*}[tb]
\centering
\includegraphics[width=0.825\textwidth]{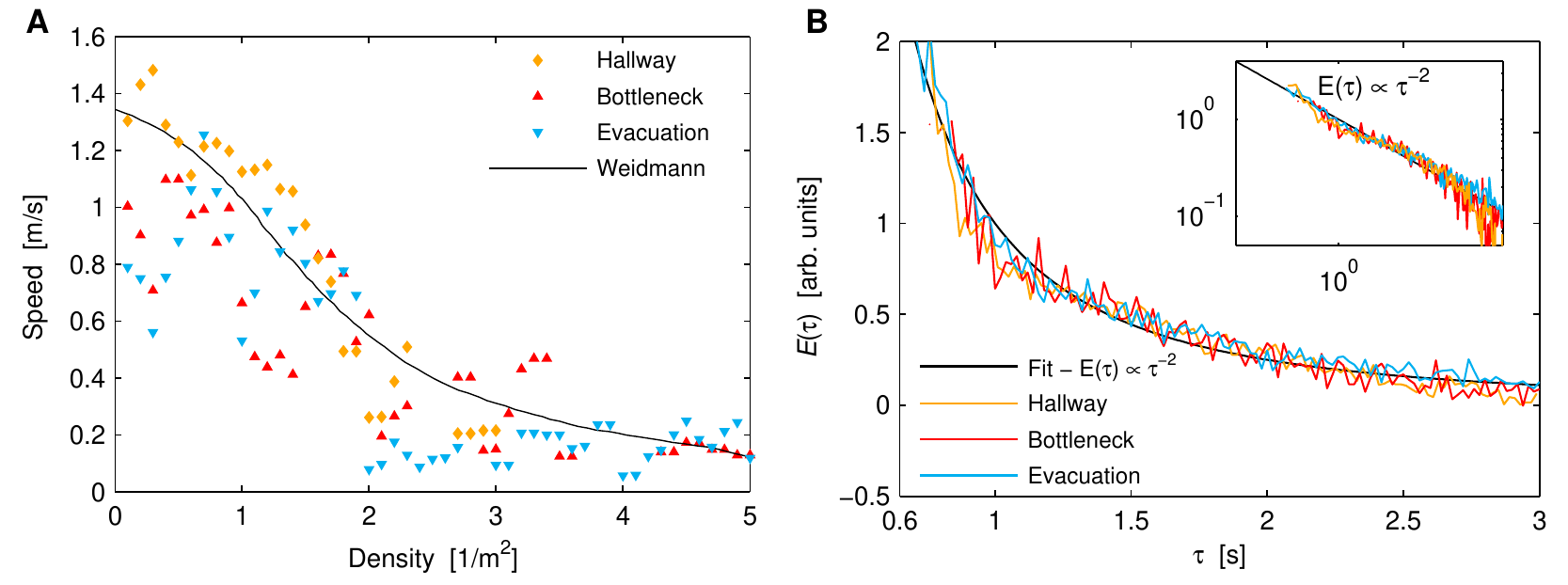}
\caption{
Analysis of simulations generated using the interaction force described in \equationref{eq:eqS2}. 
(A) Speed-density relation seen in various simulations. The results are obtained using the generalized definitions of flow, speed and density of Edie \cite{Edie63} by clustering the data into bins of $0.1\,$ppl/$\mathrm{m}^2$. The corresponding {\it fundamental diagram} is compared to measurements of Weidmann \cite{weidmann}. 
(B) The trajectories of simulated pedestrians reveal the same power-law interaction that we identify from real human crowd data (see Fig.~2 of the main text). 
}
\label{fig:figS4}
\end{figure*}

Overall, as can be seen in Fig.~3 of the main text, the derived force law described in \equationref{eq:eqS2} is able to reproduce a wide variety of collective phenomena. 
We also used the generalized definitions of flow, speed, and density suggested by Edie \cite{Edie63} to measure the density-dependent behavior that the agents exhibit in several of the simulations described above. Each simulation area was divided into two-dimensional cells measuring $0.5\,$m $\times$  $0.5\,$m and for each cell we determined its average density and speed over $4\,$s intervals. 
Figure~\ref{fig:figS4}A shows the corresponding fundamental diagram 
(here, the bottleneck samples refer to a $1\,$m wide bottleneck). 
Our results are in good agreement with the fundamental relation of speed, flow and density of real humans, as reported by Weidmann \cite{weidmann}.  

\begin{figure*}[htb]
\includegraphics[width=.775\textwidth]{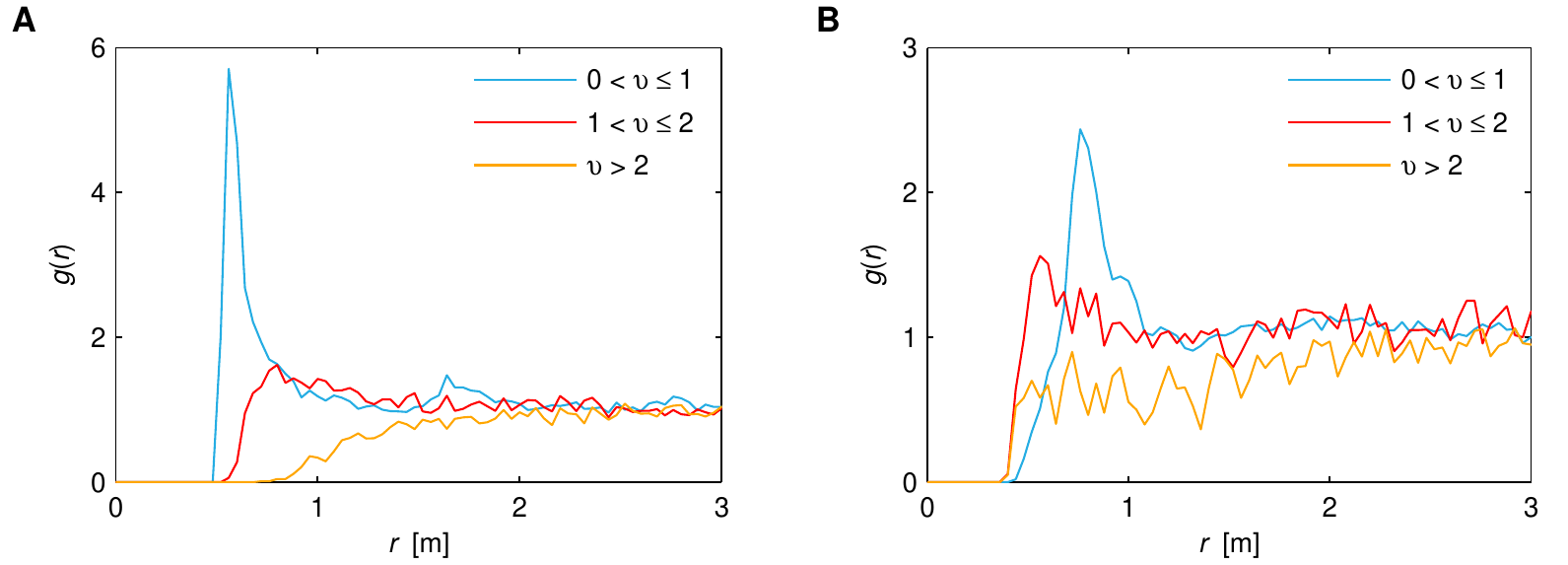}
\caption{The pair distribution function $g(r)$ in the hallway simulation generated using (A) the interaction force described in \equationref{eq:eqS2}, and (B) the distance-dependent force described in \cite{HFV00}. The variable $v$ denotes the rate of approach between pedestrians, and is measured in units of m/s. 
See also Fig.~1c of the main text.
}
\label{fig:figS5}
\end{figure*}

\begin{figure*}[htb]
\includegraphics[width=1\textwidth]{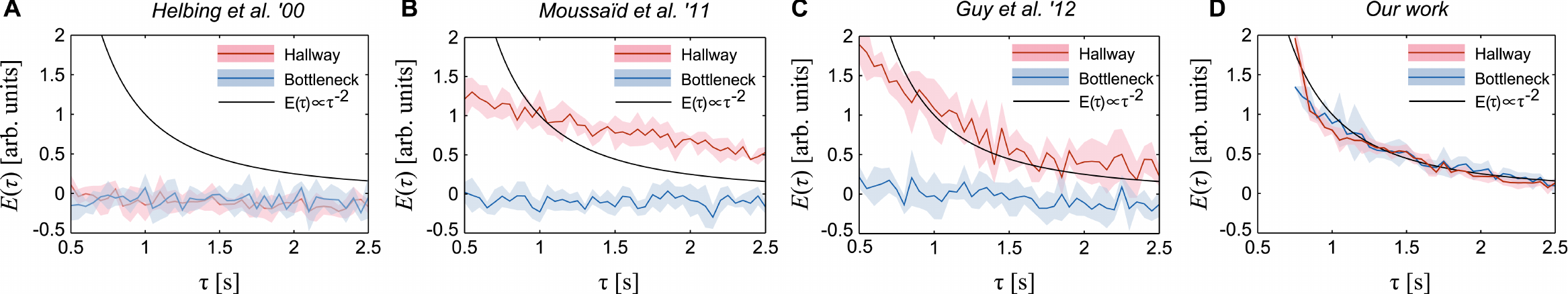}
\caption{
The inferred interaction energy $E(\tau) \propto \ln(1/g(\tau))$ as a function of time-to-collision ($\tau$) for simulations obtained using 
(A) the distance-dependent model described in \cite{HFV00}, 
(B) the behavioral heuristics model proposed in \cite{MHT11}, 
(C) the least-effort model proposed in \cite{GCL+12}, and
(D) our derived anticipatory force model. 
In all figures, colored lines indicate average energy values every 0.05$\,$s and shaded regions denote $\pm$ one standard deviation.
}
\label{fig:figS6}
\end{figure*}

Our derived force model is also sufficient to reproduce, qualitatively, the behavior of the spatial pair distribution function $g(r)$. 
This is shown in \figref{fig:figS5}A, where simulation results for $g(r)$ are plotted for different values of the pedestrian rate of approach $v$.  The strong dependence of $g(r)$ on $v$ is consistent with what we observe in the \emph{Outdoor} dataset (Fig.~1c). The peak in $g(r)$ for small $v$ can be mainly attributed to spontaneous lane formation between pedestrians walking in the same direction. 
Simulation approaches based on distance-dependent forces show somewhat different behavior of $g(r)$, as shown in \figref{fig:figS4}B, with a weaker dependence on $v$.  Such distance-dependent force simulations also fail to show a strong dependence of $g$ on the time-to-collision (as depicted in \figref{fig:figS6}A). 

Importantly, in addition to reproducing known crowd phenomena and flows, our simulations also reproduce the empirical behavior of $g(\tau)$ described in this Letter. 
Figure~\ref{fig:figS4}B shows that the inferred pedestrian interaction energy $E(\tau) \propto \ln(1/g(\tau))$ in the hallway, bottleneck ($2.5\,$m wide), and evacuation simulations closely follows the inverse quadratic power law. In contrast, existing anticipatory models of pedestrian behavior, such as the ones proposed in \cite{GCL+12,MHT11}, do not consistently capture this law.  
This is shown explicitly in \figref{fig:figS6} for the hallway and bottleneck scenarios.

\section{Note}
C++ and Python implementation of the anticipatory force model is available at \href{http://motion.cs.umn.edu/PowerLaw/}{http://motion.cs.umn.edu/PowerLaw/}, along with videos demonstrating simulation results. We also provide links to the data used to derive the power law of human interactions. 

\bibliographystyle{apsrev4-1}
\bibliography{bibliography}